\documentclass[review,number,sort&compress]{elsarticle}
\usepackage{lineno}

\usepackage{amssymb}

\begin{document}

\begin{frontmatter}



\title{Monte Carlo simulation of a very high resolution thermal neutron detector composed of glass scintillator microfibers}

\author[aff1]{Yushou Song}

\author[aff2]{Joseph Conner}
\author[aff2]{Xiaodong Zhang}
\author[aff2]{Jason P. Hayward\corref{crsautor}}
\cortext[crsautor]{Corresponding author, jhayward@utk.edu}

\address[aff1]{Key Discipline Laboratory of Nuclear Safety and Simulation Technology, Harbin Engineering University, Harbin 150001, China}
\address[aff2]{Nuclear Engineering Department, University of Tennessee, Knoxville TN 37996, USA}

\begin{abstract}
In order to develop a high spatial resolution (micron level) thermal neutron detector, a detector assembly composed of cerium doped lithium glass microfibers, each with a diameter of 1\,$\mu$m, is proposed, where the neutron absorption location is reconstructed from the observed charged particle products that result from neutron absorption. To suppress the cross talk of the scintillation light, each scintillating fiber is surrounded by air-filled glass capillaries with the same diameter as the fiber. This pattern is repeated to form a bulk microfiber detector. On one end, the surface of the detector is painted with a thin optical reflector to increase the light collection efficiency at the other end. Then the scintillation light emitted by any neutron interaction is transmitted to one end, magnified, and recorded by an intensified CCD camera. A simulation based on the Geant4 toolkit was developed to model this detector. All the relevant physics processes including neutron interaction, scintillation, and optical boundary behaviors are si\-mulated. This simulation was first validated through measurements of neutron response from lithium glass cylinders. With good expected light collection, an algorithm based upon the features inherent to alpha and triton particle tracks is proposed to reconstruct the neutron reaction position in the glass fiber array. Given a 1\,$\mu$m fiber diameter and 0.1\,mm detector thickness, the neutron spatial resolution is expected to reach $\sigma\sim 1\, \mu$m with a Gaussian fit in each lateral dimension. The detection efficiency was estimated to be 3.7\% for a glass fiber assembly with thickness of 0.1\,mm. When the detector thickness increases from 0.1\,mm to 1\,mm, the position resolution is not expected to vary much, while the detection efficiency is expected to increase by about a factor of ten.
\end{abstract}


\end{frontmatter}


\section{Introduction}
\begin{figure}[htb]
\centering
\includegraphics[width=3in]{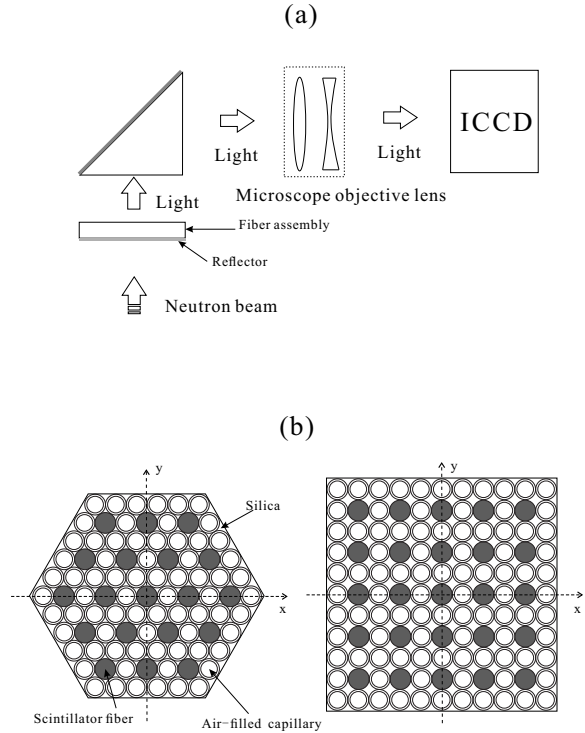}
\caption{(a)The schematic layout (not to scale) of the neutron detector system. See text for detail. (b)Both hexagonal and square microfiber assemblies are viewed in neutron beam direction.  }
\label{fig:geo}
\end{figure}

Advanced neutron sources and supporting advance\-ments in neutron instrumentation have enabled new research in advanced material characterization, energy engineering, and biological science\cite{bk:neutron_image}. At the same time, the spatial resolution of currently existing neutron detectors limits the execution of even higher-impact research \cite{jinst7.02014.2012, nc4.579.2012, jpcc116.8401.2012}. In this work, an assembly composed of cerium doped lithium glass microfibers, each with a diameter of 1\,$\mu$m, is proposed as a solution to this problem. The fiber pitch is small enough to allow one to reconstruct the neutron absorption location from the resulting charged particle products, thereby overcoming the fundamental position limitation due to finite charged particle range and the variance associated with center-of-gravity-based reconstruction. To suppress the cross talk of the scintillation light generated in any Li glass fiber, it is surrounded by air-filled glass capillaries, each having the same diameter as the fiber. This pattern is repeated to form a bulk microfiber detector (see Fig.\,\ref{fig:geo}). On one end, the surface of the detector is painted with a thin optical reflector to increase the light collection efficiency at the other end. Thus, the scintillation light emitted by any neutron interaction is transmitted to one end, bent by 90\,$^{\circ}$ by an optical mirror mounted at 45\,$^{\circ}$ with respect to the neutron beam path, magnified by a microscope objective lens, and recorded by an intensified CCD (ICCD) camera.  In this work, a Monte Carlo code based on Geant4 (version 10.0) \cite{ieeens53.270.2006} models the complete physics in the fiber assembly when a thermal neutron is incident: neutron interaction, scintillation, and optical boundary behaviors are included and treated meticulously. The optical system performance is also taken into account, although in order to show what may be possible with exellent light collection, potential sources of light loss in between the scintillator and ICCD are not modeled in detail in this work. Such detailed light loss modeling will accompany future work detailing experimental results. In this paper, we also present the results of neutron experimental measurements with a GS20 glass cylinder and a lithium glass from Nucsafe \cite{nucsafe}, which validate our Monte Carlo simulation results.

\section{Modeling of the microfiber scintillator assembly}


In this report, neutron interactions with the scintillator were handled by the data-driven high precision neutron nuclear reaction model provided by Geant4; electromagnetic processes, such as  ionization, bremsstrahlung, Coulomb scattering, were dealt with the PENELOPE \cite{g4em} code. Scintillation and optical photon processes were also considered by invoking the models in Geant4.  
The simulation data are output into the ROOT data analysis framework, which is adapted for use with Geant4 to hold data in event-based data structures. Considering in particular the optical photon portion of the simulation, there is a large magnitude of tracks to be processed. In order to avoid the need to allocate and free memory frequently, a pre-allocating memory technique supplied by ROOT was used.

\subsection{Detector modeling}
Microfiber assemblies are fabricated by a partner at the Optoelectronics Research Center at the University of Southhampton. A monolithic lithium glass rod is first drawn into rods, then the rods are packed together with air-filled capillaries and drawn, and the draw process iterated until the desired microfiber dimensions are achieved. Two possible assembly configurations are given in Fig.\,\ref{fig:geo}, where air capillaries are used to optically isolate neighboring glass scintillator fibers. One is a hexagonal pattern and the other is a square pattern (shown in Fig.\,\ref{fig:geo}(b)). The air-filled capillary has the same dimensions as the scintillator fiber, and the wall thickness is around 5\% of its diameter. The space between the scintillating and capillary microfiber is filled with silica glass during the fiber drawing process.
By convention, the incident beam direction is selected as the z axis in the Geant4 simulation. The simulated detector is placed with its axial center aligned with the z axis, and the origin coincidences with its front surface.
To save computer memory, a detector assembly of limited extent was modeled, which is large enough in extent to ensure that the charged particles produced by neutron interactions within the certain scintillator fiber (the central one) do not escape.  For a parallel neutron beam bombarding on the central fiber, an appropriate number of scintillator fibers in diameter direction $n_{fx}$ that are pack together with air capillaries in the simulation to satisfy this requirement is 150.

In this simulation, the optical system is simplified into a single photon collecting surface which is plac\-ed immediately next to the exit surface of the fiber assembly so that all exiting light is collected.  Thus, the simulated light collected in this work represents a best case scenario.

The cerium (Ce$^{3+}$) doped lithium glass scintillator supplied by Nucsafe has similar composition and optical properties as the commercial products GS20. Therefore, the properties of the scintillator fibers are set according to GS20, as specified in Table\,\ref{tab1} \cite{nima546.180.2005, saintgobain, nn17.16.2006}. To make the simulation more conservative the light yield is set as GS20's (6,000\,photon/neutron listed in the table), although Nucsafe glass is brighter according to our experimental test, which will be described in the Section \ref{sec:valid}.
In the lithium glass a neutron is captured by the reaction:
\begin{equation}\label{Li6}
  ^6\mbox{Li}+^1\!\mbox{n} \rightarrow ^4\!\mbox{He} + ^3\!\mbox{H} + 4.78 \mbox{MeV}.
\end{equation}
The scintillation light is created by the energy deposition of the alpha with kinetic energy of $E_{k\alpha}=2.05$ MeV and triton with kinetic energy of $E_{kt}=2.73$ MeV in the lithium glass. Given light output and alpha/beta ratio, the light output of the alpha $L_{\alpha}$ is inferred. Assuming the light output does not vary in the glass scintillator, the light yield per thermal neutron $Y_n$ is expressed as
\begin{equation}\label{lightn}
    Y_n=L_{\alpha}E_{k\alpha}+ L_{t}E_{kt}
\end{equation}
where $L_{t}$ is the light output due to the slowing down of the triton. Based on equation (\ref{lightn}) $L_{t}$ is inferred. The scintillation light spectrum adopted is given by \cite{saintgobain}. The scintillator is generally transparent to its emitted light \cite{nn17.16.2006} and the detector assembly is very thin, so we neglected the absorption of the light in the scintillator itself. The refractive index of the scintillator at the maximum light wavelength was implemented for the entire spectrum. The refraction of scintillation light in the silica and air for different wavelengths comes from the database given by  Polyanskiy~\cite{smeq}.
\begin{table}
\scriptsize
\renewcommand{\arraystretch}{1.3}
\caption{The properties of lithium glass scintillator GS20.}
\label{tab1}
\centering
\begin{tabular}{l l l}
\hline
Composition by weight\cite{nima546.180.2005} & SiO$_{\mbox{2}}$ & 56\% \\
                      & MgO & 4\% \\
                      &Al$_{\mbox{2}}$O$_{\mbox{3}}$ & 18\% \\
                      &Ce$_{\mbox{2}}$O$_{\mbox{3}}$ &3\% \\
                      &Li$_{\mbox{2}}$O & 18\% \\
                      &$^{\mbox{6}}$Li concentration & 95\% \\
\hline
Decay time (ns) \cite{saintgobain} & fast component  & 18 \\
                & slow component   & 57 \\
\hline
Density(g/cm$^{\mbox{3}}$) \cite{nima546.180.2005} & &  2.5 \\
\hline
Light output relative to anthracene \cite{nima546.180.2005} & & 20\%\\
\hline
Photon yield per thermal neutron \cite{nn17.16.2006} && 6000 \\
\hline
Alpha/beta ratio \cite{saintgobain} & & 0.23\\
\hline
Refractive index \cite{nima546.180.2005} & & 1.55\\
\hline
\end{tabular}
\end{table}

\section{Algorithm for neutron absorption reaction estimation}
 As incident neutron energy is far less than the nuclear reaction $Q$ value, the energetic alpha and triton move back-to-back and form a linear track. Therefore, scintillation photons are emitted along this track. Because the refractive indexes of air, the glass capillary and the silica filled among scintillator fibers and capillaries are relatively smaller than that of the scintillator fiber, total internal reflection occurs for a substantial portion of scintillation photons in the fiber. The positions of collected scintillation photons form a distribution along the ion (alpha and triton) track. This makes us be able to reconstruct the ion track as a line through Hough transformation method\,\cite{pr22.697.1989}. 
 The neutron reaction position reconstruction is implemented by the following three steps: track determination, alpha segment identification and estimation of the neutron absorption location.

\subsection{Track determination}
\begin{figure}[!hbt]%
\centering
\includegraphics[width=3in]{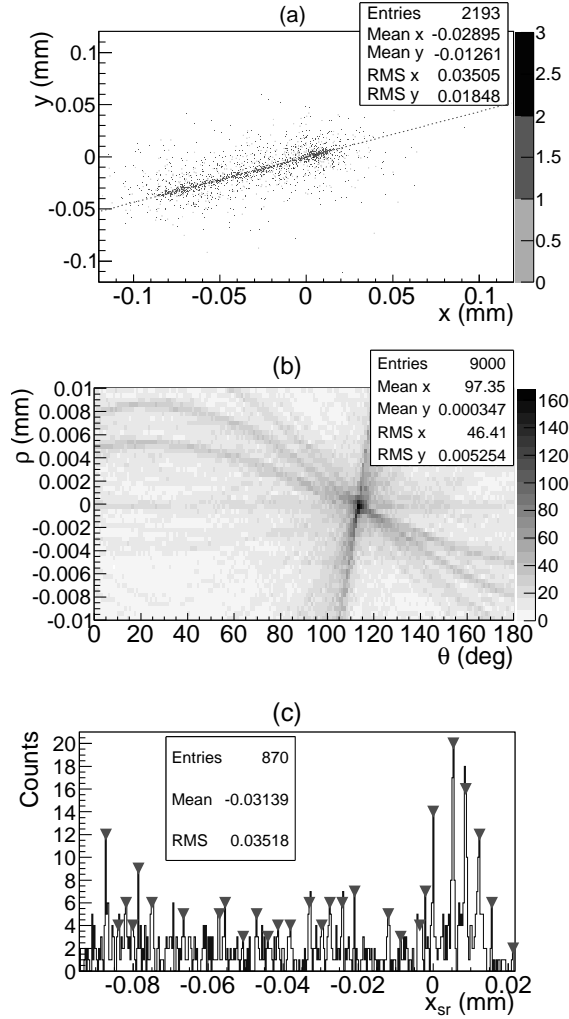}
\caption{The Monte Carlo simulation results. (a) A two-dimensional light intensity spectrum formed by a neutron absorption event, where $\delta_{pix}= 0.1\,\mu$m. The dotted line is the track line reconstructed by the Hough transformation algorithm. (b)The corresponding light intensity spectrum of (a) in Hough space. (c)The one-dimensional light intensity spectrum corresponding to (a). The inverted triangles mark the peak positions corresponding to scintillator fibers. }
\label{fig:algo}
\end{figure}

\begin{figure}[!hbt]%
\centering
\includegraphics[width=3in]{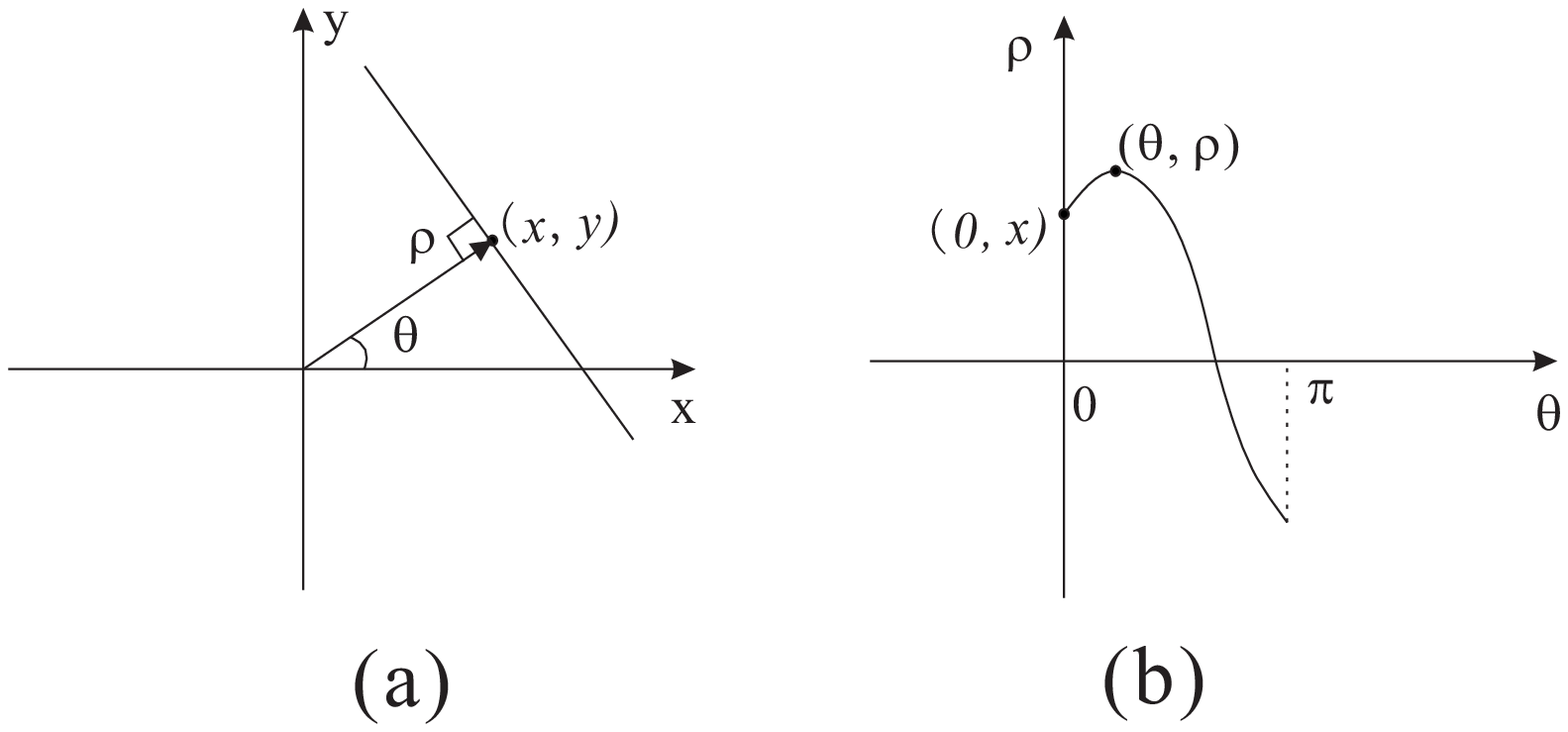}
\caption{Schematics of the Hough transformations. (a)A line going through $(x, y)$ is determined by $(\theta, \rho)$. (b) In Hough space, each point on the sinusoid stands for a line going through $(x, y)$. }
\label{fig:trans}
\end{figure}

The scintillation photons escaping from the detector assembly pass through the microscope objective lens and are detected by the ICCD, which gives a pixelated charge image recording the light intensity from the scintillator in two dimensions. The optical processes out of the detector assembly are not considered in very details.
The light intensity exiting the surface of the assembly is substituted to the charge collected on the ICCD surface.
To simulate the ICCD pixel size we introduce an equivalent pixel size $\delta_{pix}$, whose quantity is determined by the ICCD pixel $\delta_{ICCD}$ and the magnification of the microscope objective lens $f_{mag}$ according to $\delta_{pix} = \delta_{ICCD} / f_{mag}$.
Setting $\delta_{pix}= 0.1\,\mu$m, the two-dimensional light intensity spectrum is shown in Fig.\,\ref{fig:algo}(a), where the light intensity is indicated by the photon counts in a pixel.
A nonzero pixel $(x, y)$ in the light intensity spectrum becomes a weighted sinusoid after applying the Hough transformation\,\cite{pr22.697.1989} given by
\begin{equation}\label{eq:hough}
    \rho=x\cos\theta + y\sin \theta ,
\end{equation}
where $\rho$ is the algebraic distance between the origin and a line going through $(x, y)$, and $\theta$ is the angle of the vector starting from origin $(0, 0)$ and orthogonal to the line (see Fig.\,\ref{fig:trans}). When $\theta$ traverses $[0, 180^{\circ})$ all the lines going through the point $(x, y)$ are presented, with $\rho $ in the range of $[-\sqrt{x^2+y^2},$ $\sqrt{x^2+y^2}]$.
Transforming all the pixels in Fig.\,\ref{fig:algo}(a), a $(\theta, \rho)$ map forms in Hough space shown in Fig.\,\ref{fig:algo}(b), where each point stands for a line. By a maximum searching algorithm the line $(\theta_0, \rho_0)$ corresponding to the charged particle track is correctly identified.

\subsection{Alpha segment identification}
Next, the two-dimensional light intensity spectrum is projected to the reconstructed track line. 
The collimated (total internally reflected) light in the fibers is found on or around the reconstructed track and the uncollimated light populates the pixels that are further away from the track. 
Therefore, the farther a pixel is from the track line, the more likely that it contains contributions from noise alone. To improve the signal-to-noise ratio, the pixels projected to the reconstructed track line are limited by a boundary value $b_{p}$, the distance from the reconstructed track. Its estimation depends on the diameter of the fibers. In practice, $b_{p}$ may be determined by the analysis of simulation data.
Fig.\,\ref{fig:algo}(c) shows the resultant one-dimensional light intensity spectrum projected along the reconstructed track line with the boundary limitation.
The series of peaks marked with inverted triangles correspond to scintillation fibers that lie along the path corresponding to the ion track, and the height of the peak indicates ion energy deposition. The distance between the peaks at the left end and the right end gives the length $l_{ptrk}$ of the ion track projected onto the fiber assembly surface.

The neutron produced alpha and triton have different masses and kinetic energies; consequently their stopping power characteristics and ranges are also different \cite{nim135.441.1976}. From a light intensity spectrum one may identify the alpha track and the triton track, since more energy is deposited over a shorter distance for the alpha track compared to the triton track, and, thus, more light is generated and collected from these microfibers. 
In practice, two highest peaks are selected from either end of the light intensity spectrum in a range of the alpha track length. The end with larger sum value of the two peak heights is considered to be the alpha end. 
With this criterion, the accuracy to determine the alpha end from the one dimensional charge spectrum exceeds 90\% for the hexagonal pattern detector. The misjudged events mostly have very short projected track length. It does not influence the neutron absorption position estimation much, which will be explained in the following sections.

\subsection{Estimation of the neutron absorption location}
The energy deposition $\Delta E_{ion}$ of an ion within a fiber may be expressed as
\begin{equation}\label{eq:edep}
    \Delta E_{k}=\left( \frac{\mathrm{d}E_{k}}{\mathrm{d}x}\right)_{a} \Delta x ,
\end{equation}
where $(\mathrm{d}E_{k}/\mathrm{d}x)_{a}$ is the average stopping power of the ion within the fiber and $\Delta x$ is the ion path length within this fiber.
If the path lengths in each fiber are similar along the ion track, the corresponding energy deposition in different fibers may be demonstrated by the stopping power varying tendency along the ion moving direction. Starting from the neutron absorption fiber (labeled as $f_{0}$ ) the energy deposition from the alpha in each subsequent fiber (labeled as $f_{\alpha i}$, $i$=1, 2, 3...) increases roughly as $1/E_k$, then drops near the end of the track. In the opposite direction, the energy deposition from the triton in the fiber ($f_{t1}$) next to $f_{0}$ could be less or more than that in $f_{0}$, which depends on how much the alpha contributes to the energy deposition in $f_{0}$. Then it varies with the same trend as the alpha but the energy deposition within each fiber is less than that in fibers associated with the shorter alpha track. The energy deposition characteristics are interpreted by the collected light intensity spectrum as is shown in Fig.\,\ref{fig:algo}(c). Therefore, the fiber $f_{0}$ corresponds to either the minimum peak or the peak next to the minimum on the alpha track side.

Taking into account different $\Delta x$ the energy deposition varying tendency may deviate from the stopping power as the azimuthal angle $\varphi_{trk}$ of the ion track takes some values like Fig.\,\ref{eq:edep}. Thus, in order to find the neutron absorption location, the neutron absorption position estimation algorithm should be qualified to locate the scintillation fiber with minimum stopping power according to the light intensity spectrum.
For the events like what is shown in Fig.\,\ref{eq:edep} it will lead to an incorrect $f_{0}$ fiber estimation only based on the peak height of the spectrum.
To mitigate this problem, three more conditions are invoked to constrain the $f_{0}$ fiber search. i) The search algorithm should be able to recognize the alpha stopping power peak. The peak corresponding to $f_{0}$ should be estimated after going over this peak. ii)The selected minimum must have a height lower than that of the two neighboring peaks on either side. iii) The selected peak must be lower than a certain fraction of the highest peak.
The alpha particle has larger stopping power in the detector and then higher light intensity spectrum peak. To suppress the influence of statistical fluctuations and obtain the energy depositing characteristics more unambiguously, we perform a $f_{0}$ search starting on the alpha end of the light intensity spectrum.

\begin{figure}[htb]%
\centering
\includegraphics[width=3in]{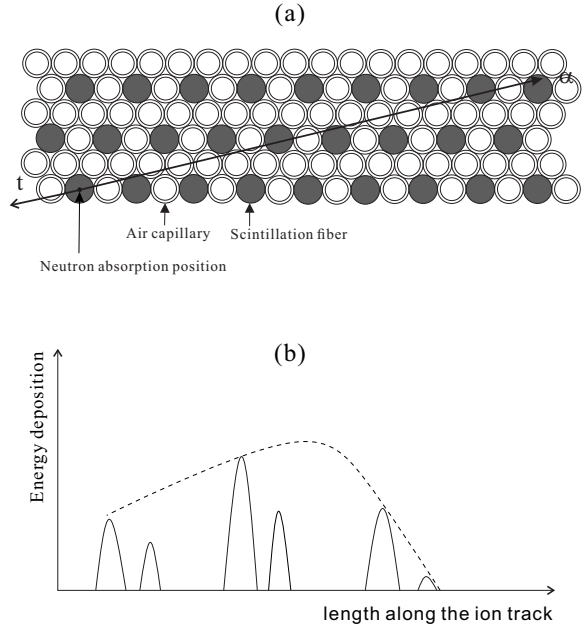}
\caption{A schematic display of how the azimuthal angle $\varphi_{trk}$ of the ion track affects the energy deposition in each fiber. (a) The ion track projection in the detector assembly (hexagonal pattern) surface plane ($xy$ plane). The alpha track  ends at the sixth scintillator fiber after it goes through five.  (b) An illustration of how alpha energy deposition varies along its track. The peaks from left to right are $f_0$, $f_{\alpha 1}$, $f_{\alpha 2}$, $f_{\alpha 3}$, $f_{\alpha 4}$ and $f_{\alpha 5}$, respectively.  $f_{\alpha 2}$ is the highest peak for this event. The dashed line indicates the expected stopping power variance of the alpha along its track. }
\label{fig:nab}
\end{figure}

The initial kinetic energies of alpha and triton ($E_{k\alpha}$ and $E_{kt}$) are the same for different neutron absorption events. Ignoring the energy deposition of the ions in the non-scintillating volumes (air-filled capillaries and silica shown in Fig.\,\ref{fig:geo}) of the detection assembly, the ratio of scintillation light yield of an alpha to the total light yield of a neutron absorption event $R_{Y\alpha n}= L_{\alpha}E_{k\alpha}/Y_n$  (see Eq.\ref{lightn}) is approximately a constant. In the light intensity spectrum, the ratio of the integral of estimated alpha track to the whole spectrum $R_{Q \alpha n}$ tends to $R_{Y \alpha n}$ if the neutron absorption position $x_{nsr}$ is estimated correctly.
Based on this fact, the light yield ratio $R_{Y\alpha n}$ may be implemented to estimate neutron absorption position $x_{nsr}$, if the algorithm fails to locate the $f_{0}$ peak by the criteria mentioned in paragraphs above.
Actual ratio of the light yield of the alpha to that of the corresponding neutron varies around $R_{Y\alpha n}$ because the ions deposit different portion of their kinetic energy in non-scintillating volumes for different events, and consequently scintillation light output fluctuates. The distribution of $R_{Y\alpha n}$ for certain detector assembly configuration can be given by the Monte Carlo simulation. It can be utilized to bound the estimation of the neutron absorption position by the $f_{0}$ peak algorithm.

If the ion track direction has a small (close to 0$^{\circ}$) or large (close to 180$^{\circ}$) polar angle $\theta_{trk}$ (the coordinate system is shown in Fig.\,\ref{fig:geo}), the track length given by ICCD is either very short or no obvious linear track is recognized. As a result, there could be very few or no peaks found from the light intensity spectrum by the peak searching process. In this case, the mean value of the whole spectrum is selected as the neutron absorption position.

\section{Validation of the simulation}\label{sec:valid}
\begin{figure}[hbt]
\centering
\includegraphics[width=3.5in]{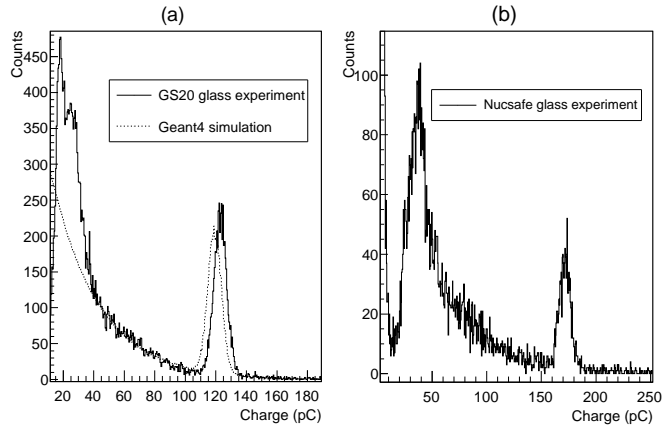}
\caption{The neutron spectra given by different glass samples. (a)The experimental charge spectrum (solid line) of a cylindrical GS20 glass sample with dimensions of 25.27\,mm (diameter) by 2\,mm (thickness) and the corresponding Geant4 simulation result (dotted line). (b)The experimental charge spectrum of a cylindrical Nucsafe glass sample with dimensions of 23.88\,mm (diameter) by 2.85\,mm (thickness). The details of the experiment are described in the text.}
\label{fig:simexp}
\end{figure}

To validate our Monte Carlo simulation we performed an experimental test with a lithium glass scintillator GS20 sample. A monolithic glass cylinder with dimensions of 25.27\,mm$\times$2\,mm (diameter$\times$thickness) was wholly wrapp\-ed with teflon tape and then black tape on top of it, only leaving one end surface coupled to the front window of a Hamamatsu photomultiplier tube (PMT) R9779 (with a bias voltage of -1400 V). The Saint Gobain BC-630 grease was used as the optical coupling medium between the glass sample and the PMT window. A $^{\mathrm{252}}$Cf source was placed 101.6\,mm away from the glass sample.
Polyethylene with a thickness of 101.6\,mm was fixed between the neutron source and glass sample as a moderator.
The $^{\mathrm{252}}$Cf source, the Polyethylene moderator, the PMT and the scintillator were placed inside of a light tight box with a blackened interior.
A DC282 digitizer from Agilent Technologies with a sampling rate of 8GSa/s was used for data acquisition. The integrated charge spectrum from the PMT is shown as the solid line in Fig.\,\ref{fig:simexp}(a).

At the same time, the experiment was modeled by our Monte Carlo code.
In the simulation a teflon tape reflection coefficient of 0.99 \cite{ieeens59.490.2012} was used. The scintillation photons arriving at the PMT R9779 photocathode were converted to photoelectrons using the PMT quantum efficiency of 0.233. The single photon response of the PMT was calibrated with a waveform generator (Agilent 33220A ) driven blue-light light emitting diode (LED) \cite{nima755.32.2014}. Under a bias of -1400 V the integrated charge corresponding to a single photon is is 88$\pm$51 fC. Implementing the calibrated single photon response with a 3.8\% Gaussian smearing to account for the electronic noise, the peak of simulated charge spectrum  (dotted line in Fig.\,\ref{fig:simexp}(a)) agrees with the experiment (histogram in Fig.\,\ref{fig:simexp}(a)) except for a shift in the peak position of $\sim $4 pC. This particular GS20 sample was slightly brighter than one might expect (according to Table \ref{tab1}).

The Nucsafe glass sample was also tested under the same experimental condition with its charge spectrum shown in Fig.\,\ref{fig:simexp}(b). According to the GS20 light output one can infer the light output of Nucsafe glass is close to 9000 photon/neutron. This proves that the modeled microfiber assembly is quite conservative.

\section{Results and discussion}
\begin{figure}[hbt]
\centering
\includegraphics[width=3.5in]{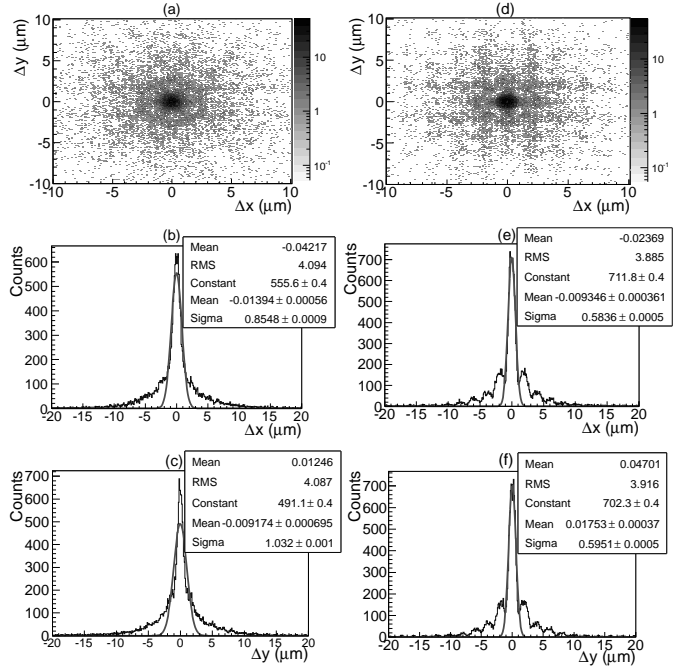}
\caption{(a) The distribution of the residual errors of reconstructed neutron absorption position in x and y directions for hexagonal configuration; (b) X projection of the residual error distribution with a Gaussian fitting; (c) Y projection of of the residual error distribution with a Gaussian fitting; (d), (e), and (f) show the corresponding plots for the square configuration.}
\label{fig:dxy}
\end{figure}

Comparing the reconstructed neutron position with the real one recorded in the simulation one may obtain the residual error $(\Delta x, \Delta y)$, which is a measure of the detector spatial resolution.The distributions of  $(\Delta x, \Delta y)$ for both hexagonal and square configurations are displayed in Fig.\,\ref{fig:dxy}. With a Gaussian fitting $\sigma _{gaus\Delta x}$ and $\sigma _{gaus\Delta y}$ reach $\sim$1\,$\mu$m for the hexagonal assembly and $\sim$0.6\,$\mu$m for square assembly.

A neutron absorption position is estimated based on the ion track projection in the xy plane (see the coordinate system in Fig.\,\ref{fig:geo}). Therefore, the residual error $(\Delta x, \Delta y)$ depends much on the ion track azimuthal angle $\varphi_{trk}$ (taking triton momentum direction as the positive track direction). The absorption positions of dominant portion of neutrons are estimated within or around the right fibers. If the algorithm fails to select the location of a fiber where neutron absorption happens, it most probably selects a fiber next to the correct one, or, with lower probability, adjacent to this one in the direction of $\varphi_{trk}$. That is why there is some clue of the fiber configuration pattern in the $\Delta x$ vs. $\Delta y$ plots. The projection in the x direction of a hexagonal fiber configuration is different from that in y direction, therefore, the $\Delta x$ and $\Delta y$ distributions are also different. However, it is symmetrical for a square fiber configuration rotating 90\,$^{\circ}$. So, the corresponding $\Delta x$ and $\Delta y$ distributions are almost the same for square configuration.

\begin{figure}[htb]
\centering
\includegraphics[width=2.5in]{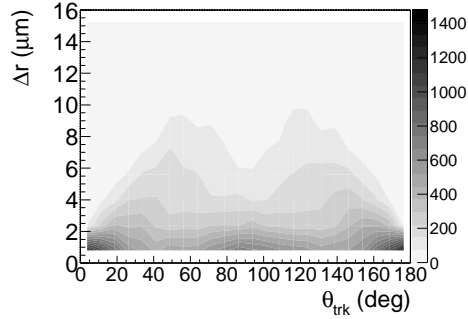}
\caption{The correlation between the deviation of estimated neutron positions $\Delta r$  and the polar angles $\theta _{trk}$ of the ion tracks. The z axis is the ion track counts per unit solid angle $\mathrm{d}N/\mathrm{d}\Omega(\theta _{trk})$ in direction of $\theta _{trk}$, where arbitrary units are used. }
\label{fig:drtheta}
\end{figure}

The deviation of the estimated neutron absorption position $\Delta r=\sqrt{\Delta x^2 +\Delta y ^2}$ also has a correlation with the polar angle $\theta_{trk}$ of an ion track as is shown in Fig.\,\ref{fig:drtheta}. The contour in the figure interprets the ion track counts per unit solid angle $\mathrm{d}N/\mathrm{d}\Omega(\theta _{trk})$, where arbitrary unit is used. For events with polar angles $\theta_{trk}$ close to 0\,$^{\circ}$ or 180\,$^{\circ}$, the ion tracks have very small intersection angles with or parallel to the fiber axis, and the track lengths formed on ICCD are very short. All the neutron absorption position are estimated by the algorithm with low uncertainty. With the increasing of the intersection angle between the ion tracks and fiber axis, more events with larger deviation $\Delta r$ appears. This tendency continues until the intersection angles increase to 50\,$^{\circ}$ $\sim$ 60\,$^{\circ}$ because the algorithm can do better in neutron absorption position estimation with the increase of the projection length of the ion track on the assembly surface, and consequently more fibers passed through and more peaks are formed in the one-dimensional light intensity spectrum.

\begin{figure}[htb]
\centering
\includegraphics[width=3.5in]{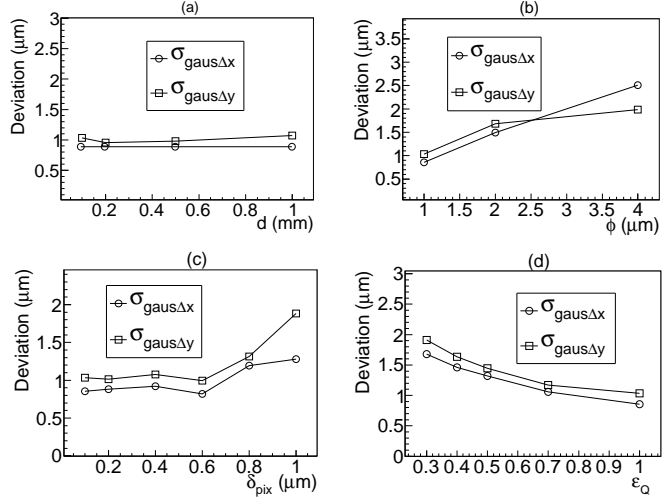}
\caption{The simulation results for hexagonal geometry. (a) The dependence of the deviation of estimated neutron absorption position on the thickness $d$ of the detector (with fiber diameter $\phi = 1 \mu$m); (b) The dependence of deviation of estimated neutron absorption position on the fiber diameter $\phi$ of the detector (with detector thickness $d=1$\,mm); (c) The dependence of deviation of estimated neutron absorption position on the spatial resolution $\delta_{pix}$ of the optical photon collection system. (d) The dependence of deviation of estimated neutron absorption position on the ICCD quantum efficiency $\varepsilon_{Q}$. In (c) and (d) the  fiber diameter $\phi = 1 \mu$m and detector thickness $d=1$\,mm.}
\label{fig:vary}
\end{figure}

To aid in detector design and fabrication, we also investigated how expected spatial resolution depends upon certain design parameters(see Fig.\,\ref{fig:vary}), taking the hexagonal configuration as an example.
When the detector thickness increases from 0.1\,mm to 1\,mm, $\sigma _{gaus\Delta x}$ and $\sigma _{gaus\Delta y}$ do not vary much. However, the portion of the events under the Gaussian fit curve changes from 66\% to 56\%. This means that the portion of the fine events estimated to the correct fiber decreases. When the fiber diameter changes from 1\,$\mu$m to 4\,$\mu$m, $\sigma _{gaus\Delta x}$ and $\sigma _{gaus\Delta y}$ become worse with the increasing of the fiber diameter (see Fig.\,\ref{fig:vary}(b)). While the resolution is not expected to increase as quickly as the fiber diameter because photon statistics improve as the fiber diameter increases.

One more factor that may influence the position resolution is the detector assembly area corresponding to an ICCD pixel $\delta_{pix}$, which is decided by the design and geometrical arrangement of the microscope objective lens. From Fig.\,\ref{fig:vary}(c), one may conclude that $\delta_{pix}$=0.6\,$\mu$m is enough for a fiber diameter of 1 micron. Now the smallest available ICCD (model iStar 340T) \cite{andor} pixel dimension $\delta_{ICCD}$ has reached 13.5\,$\mu$m (with a position resolution $\sim$27\,$\mu$m), that means that a microscope with a magnification of 45 satisfies the detection system requirement.
The quantum efficiency of an ICCD reduces the statistics of charge spectrum, which is another factor affects the position resolution of the detector (see Fig.\,\ref{fig:vary}(d)). For the ICCD model iStar 340T with a quantum efficiency $\varepsilon_{Q}=50\%$,  $\sigma_{gaus\Delta x}=1.3 \mu $m and $\sigma_{gaus\Delta y}=1.4 \mu $m. Even the quantum efficiency drop to 0.3, $\sigma_{gaus\Delta x}$ and $\sigma_{gaus\Delta y}$ are still below 2$\mu$m.

With detector thickness of 0.1\,mm and a scintillator fiber diameter of 1$\,\mu$m , the intrinsic detection efficiency is estimated to be 3.7\% by our simulation, which agrees with the analytic calculation that considers $^\mathrm{6}$Li neutron cross section and scintillator geometry. Increasing the detector thickness from 0.1\,mm to 1\,mm, the efficiency will have a ten times of rise linearly. 

\section{Conclusion and outlook}
In this paper we proposed a thermal neutron detector with very high position resolution. With the scintillator fiber and air capillary arrangement described, the scintillator light produced in scintillating fibers is expected to be effectively isolated through total internal reflection.
To evaluate the performance of either hexagonal or square detector assemblies, a simulation code based on the Monte Carlo toolkit Geant4 was developed.
To reconstruct the neutron position from experimental data, an algorithm uses the information collected from individual neutron-produced alpha and triton trajectories. This algorithm includes three procedures: track direction determination, alpha segment identification and estimation of the neutron absorption location. For fiber diameter $\phi$=1\,$\mu$m, the expected position resolution ($\sigma _{gaus\Delta x}$ and $\sigma _{gaus\Delta y}$) is around 1\,$\mu$m for the hexagonal assembly pattern and about 0.6\,$\mu$m for the square assembly pattern. If the position resolution requirement is not as firm, an assembly with larger diameter fiber (e.g. $\phi$=4\,$\mu$m) still has a resolution smaller than the fiber cross section, which has lower requirements for the photon collecting system (microscope and ICCD). Future work will report on testing of fiber assemblies at a neutron beamline.

\section*{Acknowledgment}
This material is based upon work supported by the U.S. Department of Energy, Office of Science, Office of Basic Energy Sciences, under Early Career Award Number DE-SC0010314, and the National Natural Science Foundation of China(Grant No.11205036). The authors also would like to thank Joshua Cates, now at Stanford, for the work involved with his first simulations in 2D of charge particle tracks and light collection in Li glass assemblies.





\section*{References}

\bibliographystyle{elsarticle-num}
\bibliography{./nfiber}

%
%
%

\end{document}